\documentclass[11pt]{article}
\usepackage{a4wide}
\usepackage{amssymb}
\usepackage{xcolor}
\usepackage{hyperref}
\usepackage{amsmath}

\newcommand{\bea}{\begin{eqnarray}}
\newcommand{\eea}{\end{eqnarray}}

\def\nn{\nonumber}
\def\d{{\rm d}}

\begin{document}
{\renewcommand{\thefootnote}{\fnsymbol{footnote}}
\begin{center}
{\LARGE  Non Bunch Davies group coherent states, and their quantum signatures in CMB observables}\\
\vspace{1.5em}
Pietro Don\`a$^{1,2}$\footnote{e-mail address: {\tt pietro\_dona@fudan.edu.cn}}
and Antonino Marcian\`o$^1$\footnote{e-mail address: {\tt marciano@fudan.edu.cn}}\\
\vspace{0.5em}
$^1$ Department of Physics \& Center for Field Theory and Particle Physics,
Fudan University,
200433 Shanghai, China\\
\vspace{0.5em}
$^2$ CPT, Aix-Marseille Universit\'e, Universit\'e de Toulon, CNRS, F-13288 Marseille, France.\\
\vspace{1.5em}
\end{center}

\setcounter{footnote}{0}

\begin{abstract}
\noindent
We approach the problem of finding generalized states for matter fields in a de Sitter universe, moving from a group theoretical point of view. This has profound consequences for cosmological perturbations during inflation, and for other CMB observables. In a systematic derivation, we find all the allowed generalities that such a state may have, within the requirements of coherent states. Furthermore, we show their relation to the more familiar excited initial states, often used in inflation.
\end{abstract}

\section{Introduction}\label{1}
Matter fields, which must be described as fundamentally quantum objects, source the Einstein equations for the evolution of space-time backgrounds, entering as (observable) expectation values. There are thousands of examples in which quantum matter fields have been deployed to cosmology, and specifically in the computation of the expectation values of their energy-momentum tensor (for a detailed presentation of the subject see e.g. \cite{Birrell:1982ix} and references therein). An immediate question, which arises naturally at this stage, is as follows: which is the quantum state that must be taken into account in performing such an expectation value? The choice, of course, depends on the physical framework under scrutiny, and on the boundary conditions that are imposed on the initial Cauchy surface.

For instance, in most part of the literature on inflation, the vacuum state of the quantum matter content under study is selected as the standard reference. More specifically for inflation, the initial state one usually refers to is the vacuum state of the bosonic Hilbert space on the de Sitter background \cite{Bunch:1978yq}. This state is pure (and indeed, Gaussian), but it can be also regarded as a thermal state, with zero temperature, which makes it suitable to the description of a universe that undergoes inflation while starting with initial zero entropy \cite{Ion}. Entropy then grows at the end of inflation, thanks to the decay of some bosonic (usually, scalar inflaton) field, which was driving inflation. Nonetheless, it might be difficult to reconcile this picture with the thermal history of the universe \cite{Kolb}, since we expect at the onset of inflation the universe to be dominated by a gluon-fermion plasma with temperature at the GUT energy scale. What we would naturally imagine then is that picking up some macroscopic quantum state of matter, which is able to account for this condensate phase that would be a better fit as the initial state for the physical situation under examination.

On the other side, what has been then clarified in more recently in the literature \cite{Vilenkin:1987kf, Brandenberger:2000wr, Kaloper:2002cs, Holman:2007na, Ashoorioon:2010xg, Ashoorioon:2013eia, Ashish, Benedict}, is that the choice of the initial state can actually carry profound consequences for the fate of the models of early cosmology. Indeed predictions that can be derived about some main observables of the cosmological microwave background radiation (CMBR) heavily rely on which quantum state is finally selected. This is true not only for what concerns the estimate of the non-Gaussianities, but also applies  to derivation of the power spectra of CMBR, like the spectrum of scalar cosmological perturbations, and of the tensor to scalar ratio parameter. Somehow, the choice of the quantum matter state to be specified on the initial Cauchy surface must be considered to be part of the choice of the initial conditions. In a quantum many-body system perspective, like the one outlined in Ref.~\cite{Dona:2016fip}, the states to be considered in cosmology should rather correspond to some specific coherent states, which is indeed accounting for a choice of initial conditions and is describing a condensate phase for the matter system.

No matter if developed within the context of late or early time cosmology, if we aim at recasting this kind of analyses within the wider theoretical framework of the modern theory of quantum measurement \cite{WiMi}, we shall recognize that both the quantum system and the quantum probe shall play a crucial role in the measurement results, and thus both shall be taken into account in the analysis. This is exactly our perspective while addressing cosmological consequences that arise form inflation, for which the choice of initial states that are ``non Bunch-Davies'' --- if we want to use the jargon adopted in the literature --- becomes relevant.

Afterwards, the gist of the analysis here developed reckons on the truly quantum nature of the matter system on which we are focusing, and on the consideration of its very quantum (Hilbert space's) states. Therefore, this perspective naturally calls for the analysis of peculiar quantum features that can be deepened for the matter system. These features are present within the investigations of the many-body systems, which indeed triggered the analysis of cosmological coherent states reported in Ref.~ \cite{Dona:2016fip}, while seeking macroscopic quantum states that enable recovering cosmological backgrounds from the Einstein equations. The presence of quantum entanglement long-range effects in cosmology are a byproduct of this perspective. Although very fashionable, at the present time it is practically impossible to trace back to the CMB the detection of primordial entanglement of the Universe \cite{Martin-Martinez:2014gra}. Light strongly interact with matter, and after inflation, too long time has been occurred until recombination for entanglement to survive. Neutrinos, which decoupled from matter very rapidly, might represent a more reliable source for detecting entanglement, but our knowledge of cosmic neutrino background is still very limited.

Nonetheless, the framework we are accounting for in this paper can be also deployed to late time cosmology too, so that the current expansion of the Universe can be now linked to the detection of entanglement-harvesting signature \cite{Martin-Martinez:2014gra}. This is actually a proposal (see Ref.~\cite{VM}), which elaborates on the discovery by Reznik that field local detectors can harvest entanglement \cite{R1,R2} and thus makes possible to distinguish experimentally among conformal (see e.g. Ref.~\cite{VM}) or minimal coupling of matter fields with to the metric, also during inflation \cite{N}.

Given these issues, which remain as open perspectives for forthcoming speculations, in this paper we will focus on a constructive way to find generalized states for matter fields in a de Sitter universe, deploying a group theoretical procedure. Thus, we will be more concerned about the application of this framework to inflation, with particular focus on primordial cosmological perturbations. We will then analyze the consequences for the nearly scale-invariance of the CMB spectrum of scalar gravitational perturbations, and at the same time focus on the other relevant observables, including the estimate of the parameters of non-Gaussianity, and the possibility of describing modes coupling in this set up.

The plan of the paper is the following. In Sec-\ref{2} we review the solutions of the equation of motion and recover the mode functions for a minimally coupled real scalar field on de Sitter spacetime. We then introduce the Bunch-Davies vacuum state, and comment on its Bogolubov transformation being a coherent state that shows thermal properties \cite{MartinMartinez:2012sg}, and can be refer to as a prototype of non Bunch-Davies state. In Sec-\ref{3}, we focus on bosonic statistics, and generalize the group theoretical construction of non Bunch-Davies states for de Sitter spacetime, so to account for the construction of coherent states for different Lie groups. In Sec-\ref{4}, we comment on the possible signatures that might be detected for CMB observables, specifically in the power spectrum of primordial gravitational scalar perturbations, and in the estimate for the bispectrum. Finally, concluding remarks and outlooks are contained in Sec-\ref{5}.

\section{Real scalar field on de Sitter background}\label{2}
Let us first review what is known as `non Bunch-Davies' states in the literature for scalar fields on a de Sitter (dS) background. We begin with the action for a massive scalar on dS
\bea\label{action1}
S = \frac{1}{2}\int  \d^4x \sqrt{-det(g)} \left[g^\mu\nu \nabla_\mu\varphi \nabla_\nu\varphi -m^2\varphi^2\right]\,,
\eea
where the metric is defined through the line element
\bea \nonumber
\d s^2= a^2(\eta)\left[\d\eta^2 -\delta_{ij}\d x^i\d x^j\right] = \frac{1}{H^2\eta^2}\left[\d\eta^2 -\delta_{ij}\d x^i\d x^j\right]\,,
\eea
with $\eta, a(\eta)$ and $H$ denotes the conformal time parameter, the scale factor and the Hubble parameter respectively. The greek indices run from $0$ to $3$ while the latin ones denote spatial coordinates from $1$ to $3$.
Following standard convention, we introduce the variable $\phi = a(\eta)\varphi$ in the action (\ref{action1})
\bea\label{action2}
S = \frac{1}{2}\int \d^3x \d\eta\left[\phi'^2 - \partial_i\phi\partial^i\phi - m^2_{\text{eff}}(\eta)\phi^2\right]\,,
\eea
where the effective mass is now time dependent, $m_{\text{eff}}(\eta) = \sqrt{\left(m^2/H^2-2\right)/\eta^2}$. The equation of motion for the scalar field follows from this action
\bea\label{EOM1}
\phi'' -\partial^2\phi + m^2_{\text{eff}}\phi = 0\,.
\eea
Going to the Fourier space
\bea
\phi(\vec{x}, \eta) = \int \frac{\d^3\vec{k}}{(2\pi)^{3/2}} \phi_{\vec{k}}(\eta)e^{i\vec{k}.\vec{x}}\,,
\eea
and introducing the time-dependent mode frequency $\omega^2_k(\eta) = k^2+m^2_{\text{eff}}(\eta)$, the mode equations are
\bea\label{EOM2}
\phi''_{\vec{k}} +\omega^2_k\phi_{\vec{k}} = 0\,.
\eea
Since we are dealing with a real scalar field, $\phi^*_{\vec{k}}(\eta)=\phi_{-\vec{k}}(\eta)$, the mode expansion of the scalar field can be written as
\bea\label{ModeExp}
\phi(\vec{x}, \eta) = \frac{1}{\sqrt{2}}\int \frac{\d^3\vec{k}}{(2\pi)^{3/2}} \left[ a_{\vec{k}}^- u^*_k(\eta) e^{i\vec{k}.\vec{x}} + a_{\vec{k}}^\dagger u_k(\eta) e^{-i\vec{k}.\vec{x}} \right]\,.
\eea
In the above expansion, $u_k(\eta)$ and $u^*_k(\eta)$ are two linearly independent solutions of (\ref{EOM2}). Requiring that the ladder operators satisfy the standard commutation relations, $\left[a^-_{\vec{k_1}}, a^\dagger_{\vec{k_2}}\right] = \delta^3(\vec{k_1}- \vec{k_2})$, we find that the mode functions must satisfy the normalization (Wronskian) condition
\bea\label{Normalization1}
u'_ku_k^* - u_k u_k^{*\prime}=2i\,.
\eea
For an exactly dS spacetime, these mode functions can be chosen to the ones corresponding to the Bunch-Davies (BD) vacuum state, which is defined as the one annihilated by the operator $a_{\vec{k}}^-$. There are two conditions required for choosing this preferred vacuum in dS.
\begin{enumerate}
 \item The vacuum state must respect the isometries of dS spacetime, i.e. it must be invariant under the actions of the full dS group.
 \item The deep UV behaviour of the vacuum state must approach that of the Minkowski vacuum at an appropriate rate as given by the Hadamard condition \cite{Had}.
\end{enumerate}
In other words, the physical choice for the vacuum state come from the consideration that, for modes of any given wavelength, going back to early enough times, it shall have a physical wavelength which is much smaller than the Hubble length, i.e.
\bea
\frac{H^{-1}}{k |\eta|} \ll H^{-1}\,.
\eea
Thus these modes do not feel the curvature due to gravity and behave as if they are in Minkowski space. In this case, there are two independent wave solutions to the mode equation, and one chooses the negative-frequency modes as the vacuum one, $u_k(\eta) \approx \frac{1}{\sqrt{\omega_k}}\, \exp{i \omega_k \eta}$. This is the required initial condition ($\eta\rightarrow -\infty$) on the solution to the mode equations for a real scalar field in dS, to obtain the BD vacuum, characterized by the modes
\bea
u^{\text{BD}}_k(\eta) = \sqrt{-\frac{\pi\eta}{2}} \left[J_\nu \left(-k\eta\right) - iY_\nu\left(-k\eta\right)\right]\,,
\eea
introducing a new parameter related to the mass of the scalar field, $\nu = \sqrt{\frac{9}{4}-\frac{m^2}{H^2}}$. It can be shown that the BD vacuum is a time-independent, stationary  state.

However, the above argument also shows that one can turn this argument around to have non-BD initial states for such a scalar, if cannot extend dS space to the beginning. In the context of inflation, if the dS space started at a finite time, $\eta_0$, then some of the modes may be excited at that point. it is now somewhat common to consider excited initial states for inflation which reduce to the BD one for deep UV modes (following the Hadamard condition). These states are generally the Bogolyubov rotations on the BD state and goes under the (slightly confusing) name of `non BD' initial states. (Other more general states have been considered through defining a density matrix at some initial time $\eta_0$ when inflation started.) The underlying idea here is that the most general solution to the equation (\ref{EOM2}) is given by
\bea
\label{bogolibovmode}
v_k(\eta) = \alpha_k u^{\text{BD}}_k(\eta) + \beta_k u^{\text{BD}*}_k (\eta)\,,
\eea
with the normalization condition (\ref{Normalization1}) requiring that
\bea
|\alpha_k|^2 - |\beta_k|^2 = 1\,.
\eea
The general motivation for using such excited states is the urge to parametrize our ignorance regarding pre-inflationary dynamics. (One specific example of phase transition into inflation from a different era leading to non BD states is given in \cite{VilenkinFord}.) Although these states are different from the usual BD vacuum, they are, however, still pure and indeed, even, Gaussian. In order to go beyond such Gaussian states, one approach has been to consider mixed density matrix at some initial time with non-Gaussian terms in the action \cite{Nishant}. In this paper, aim is to organize all such states, from a completely different point of view of group coherent states, in a systematic manner.

\subsection{Interpreting these as coherent states}
Instead of writing the new mode functions in terms of the old BD ones as in (\ref{bogolibovmode}), we might express a new set of creation and annihilation operators (which obey the same standard commutation relation as the BD ones), $\hat{\tilde{a}}^\pm_{\vec{k}}$, in terms of the BD ones from the relations
\bea
\label{bogolyubovladderop}
\hat{\tilde{a}}_{\vec{k}}^- &=& \alpha_k \hat{a}^-_{\vec{k}} + \beta_k \hat{a}^+_{-\vec{k}}\,,\nn\\
\hat{\tilde{a}}_{\vec{k}}^+ &=& \alpha_k^* \hat{a}^+_{\vec{k}} + \beta_k^* \hat{a}^-_{-\vec{k}}\,.
\eea
However, to interpret the new state as a coherent state different from the BD vacuum, it is instructive to write down this state as a linear combination in the old basis. The vacuum for the new $\tilde{a}$ particles, for a single mode, can be expanded as
\bea\label{newvacuum1}
|0_{\vec{k}, -\vec{k}} \rangle_{\tilde{a}} = \sum_{i,j=0}^\infty c_{ij} |i_{\vec{k}}, j_{-\vec{k}}\rangle_{\text{BD}}\,,
\eea
where the generic state $|i_{\vec{k}}, j_{-\vec{k}}\rangle_{\text{BD}}$ is an excited BD state created by acting with $i$ number of creation operators $\hat{a}^+_{\vec{k}}$ and $j$ number of creation operators $\hat{a}^+_{-\vec{k}}$ on the BD vacuum, i.e.
\bea\label{newvacuum2}
|i_{\vec{k}}, j_{-\vec{k}}\rangle_{\text{BD}} = \frac{1}{\sqrt{m!n!}} \left(\hat{a}_{\vec{k}}^+\right)^i \left(\hat{a}^+_{-\vec{k}}\right)^j |0\rangle_{\text{BD}}\,.
\eea

The arbitrary coefficients $c_{ij}$ can be calculated by using the fact that the new vacuum is annihilated by the $\hat{\tilde{a}}^-_{\vec{k}}, \hat{\tilde{a}}^-_{\vec{k}}$ operators. Using the definition of these operators in terms of the BD ones, as given in (\ref{bogolyubovladderop}), we find
\bea\label{newvacuum3}
\left(\alpha_k \hat{a}^-_{\vec{k}} + \beta_k \hat{a}^+_{-\vec{k}}\right) |0_{\vec{k}, -\vec{k}} \rangle_{\tilde{a}} &=& 0\,,\nn\\
\left(\alpha_k^* \hat{a}^+_{\vec{k}} + \beta_k^* \hat{a}^-_{-\vec{k}}\right) |0_{\vec{k}, -\vec{k}} \rangle_{\tilde{a}} &=& 0\,.
\eea
After a somewhat long but straightforward amount of algebra, we can solve for this coefficients to find that
\bea
|0_{\vec{k}, -\vec{k}} \rangle_{\tilde{a}} = \frac{1}{|\alpha_k|} \sum_i^\infty \left(\frac{\beta_k}{\alpha_k}\right)^i |i_{\vec{k}}, i_{-\vec{k}}\rangle_{\text{BD}}\,,
\eea
for each mode\footnote{To get the new vacuum state, we need to take a tensor product of each one of these states for all the modes.}. The structure of this new state immediately lets us conclude that this is a $SU(1, 1)$ group coherent state, as explained in the next section.

\section{General class of coherent states}\label{3}
The idea of coherent states in quantum system trace back almost to the origin of quantum mechanics and Schr\"oedinger in 1926 in the context of the classical motion for an harmonic oscillator. This simplest construction of a coherent state is usually associated with the Heisenberg-Weyl group, whose Lie algebra is given in terms of a harmonic oscillator creation and annihilation operators. The coherent states for arbitrary Lie group was studied by Perelomov in \cite{Perelomov}.

The construction of representations of the compact Lie algebra $SU(2)$ in terms of creation and annihilation operators is due to Schwinger in 1952~\cite{Schwinger}. The non-compact case was discussed by Barut and Fronsdal in 1965 for $SU(1, 1)$ \cite{Barut} and by Anderson, Fischer and Raczka in 1968 for $U(n, m)$ \cite{Anderson}.

By mixing the Perelomov construction and the Lie Algebra realization in terms of annihilation and creation operators we can define Lie group coherent states in the Fock space of a scalar field. While dealing with both bosonic and fermionic statistics, this construction has been applied by some of us to cosmology, for the analysis of cosmological perturbations \cite{Dona:2016fip}.

We consider a representation of the $SU(1, 1)$ algebra composed by three generators $K_\pm$ and $K_3$ with commutation relations
\begin{equation}
\label{su11}
\left[K_3, K_+\right] = K_+\ , \qquad \left[K_3, K_-\right] = -K_-\ , \qquad \left[K_+, K_-\right] = 2 K_3\ .
\end{equation}
We note that this (unitary) representation is necessarily infinite dimensional, we will therefore always consider in the following the so called discrete series representations which are the analogous of the usual $SU(2)$ ``spin'' irreducible representation.

The bosonic representation of the $SU(1,1)$ Lie algebra is realized on the the Hilbert space of two harmonic oscillators defined in terms of the creation (annihilation) operators $a^\dagger_{1,2}$ ($a_{1,2}$). The generators $K_s$ on this Hilbert space are realized by the following operators
\begin{equation}
K_+ = a_1^\dagger a_2^\dagger\ , \qquad K_- = a_1 a_2\ ,\qquad K_3 = \frac{1}{2} \left(a_1^\dagger a_1 +a_2^\dagger a_2 + 1\right) \ ,
\end{equation}
which satisfy the commutation relation \eqref{su11}. The definition of a generalized coherent state consists in the action of a general element of the group (the exponential of a linear combination of the generators) on a reference state.

In the Fock space, given two independent modes of the scalar field, we can define an $SU(1,1)$ representation, and a general group coherent state for the scalar field will be the tensor product of all of them. Moreover we will consider as a reference state the BD vacuum which transforms in the fundamental representation. Up to an irrelevant phase the class of $SU(1,1)$ coherent state under scrutiny have the following form
\begin{equation}
\left| \alpha \right> \equiv D(\alpha)\left| 0 \right> \equiv \exp \left[\int_{k_1\neq k_2} \alpha(k_1,k_2) a_{k_1}^\dagger a_{k_2}^\dagger - \alpha^*(k_1,k_2) a_{k_1} a_{k_2} \right] \left| 0 \right>\,.
\end{equation}

Looking forward for applications to cosmology, and in particular to the inflationary scenario, we may require hat this state to be homogeneous and isotropic.
Requiring that $\left| \alpha \right>$ is invariant over space translation fix the function $\alpha(k_1,k_2)$ to be proportional to a Dirac delta of the two momenta
\begin{equation}
\label{homogeneouseq}
\exp\left( i \hat{P}\cdot x\right) \left| \alpha \right> = \left| \alpha \right> \Rightarrow \alpha(k_1,k_2)=\alpha(k_1) \delta\left(k_1+k_2\right)\,.
\end{equation}

Similarly, requiring that $\left| \alpha \right>$ is invariant over space rotation fix the function $\alpha(k)$ to depend only on the modulus of the momentum $k$
\begin{equation}
\hat{R}(\theta) \left| \alpha \right> = \left| \alpha \right> \Rightarrow \alpha(k)=\alpha(|k|)\,.
\end{equation}

We can rewrite the most general homogeneous and isotropic $SU(1,1)$ group coherent state for a scalar field in the following way:
\begin{equation}
\label{cstatefinal}
\left| \alpha \right> \equiv \exp \left[\int_{k} \alpha(|k|) a_k^\dagger a_{-k}^\dagger - \alpha^*(|k|) a_k a_{-k} \right] \left| 0 \right>\,.
\end{equation}

The expectation value of any observable on a coherent state $\left| \alpha \right>$ can be easily related to its vacuum expectation value:
\begin{equation}
 \left< \alpha \right| \mathcal{O}\left(a_k , a_k^\dagger \right)\left| \alpha \right> =\left< 0 \right| \mathcal{O}\left(\tilde a_k , \tilde a_k^\dagger \right)\left| 0 \right>\,,
\end{equation}
where $\tilde{a}_k$ are the Bogoliubov transformed of $a_k$ \eqref{bogolibovmode}:
\begin{equation}
\label{bogoliubovop}
 \tilde a_k = D^\dagger(\alpha) a_k D(\alpha) = \cosh\left(\left|\alpha(|k|)\right|\right)a_k + \frac{\alpha(|k|)}{|\alpha(|k|)|} \sinh\left(\left|\alpha(|k|)\right|\right) a^\dagger_{-k}\,.
\end{equation}

\subsection{Infinite class of such states?}
Accounting for different examples of Lie algebras, in this section we analyze the representation of their generators of a Lie algebra on a Hilbert space with bosonic statistics. Our purpose is therefore to check whether is it possible to generalize the initial states for inflation considered above to any Lie group coherent states.

\subsubsection{Possible generalizations: $U(n,m)$.}
We will give first the general argument and then we will clarify the procedure by giving some explicit examples concerning the $SU(2)$, $SU(2,1)$ and $SU(2,2)$ cases.
A realization of the Lie algebra of the group $U(n, m)$ can be obtained with bilinear products of bosonic annihilation and creation operators. In particular, we can construct a basis for this algebra following the construction in \cite{Anderson}. We can then start from two independent sets of ladder operators, $a_{k_i}$ with $i=1,\ldots, n$ and $a_{q_j}$ with $j=1,\ldots, m$, and consider  
\begin{align}
\label{goodguys}
&&- a^\dagger_{k_i} a_{k_{i'}}  \quad {\rm for} \quad && i',i=1,\ldots, n\,;&& a^\dagger_{k_i} a^\dagger_{q_j}  \quad {\rm for} && \Big\{  \begin{array}{l}
i=1,\ldots, n \\
j=1,\ldots, m
\end{array}\,;\\
&& a_{q_j} a^\dagger_{q_{j'}} \quad {\rm for} && j',j=1,\ldots, m\,;&& a_{k_i} a_{q_j}  \quad {\rm for} &&  \Big\{ \begin{array}{l}
i=1,\ldots, n \\
j=1,\ldots, m
\end{array} \,.
\end{align}
We define a generalized coherent state by acting with a general element of the group on a reference state. If we insist in requiring the vacuum to be the reference vector, which is necessary to get a homogeneous state, the only operators that gives a non vanishing or non trivial contribution are the one of \eqref{goodguys}. The first consequence is that both $n\geq 1$ and $m\geq 1$ are required. The coherent state class take the explicit form
\begin{equation}
\left| \alpha \right> \equiv D(\alpha)\left| 0 \right> \equiv \exp \left[\int_{k_i,q_j} \alpha(k_i,q_j) a_{k_i}^\dagger a_{q_j}^\dagger - \alpha^*(k_i,q_j) a_{k_i} a_{q_j} \right] \left| 0 \right>\,.
\end{equation}
Similarly to the case of $SU(1,1)$, by imposing the invariance under translation we get  $p\cdot q$ equations each of them will couple one momenta $k$ with one momenta $q$. The maximum set of indipendent solution that can be found are $min(n,m)$, while all the other coefficients $\alpha$ have to be set to zero. Thus we finde
\begin{equation}
\left| \alpha \right> \equiv D(\alpha)\left| 0 \right> \equiv \exp \left[\sum_{k_i} \int_{k_i} \alpha(k_i) a_{k_i}^\dagger a_{-k_i}^\dagger - \alpha^*(k_i) a_{k_i} a_{-k_i} \right] \left| 0 \right>\,,
\end{equation}
and by making a change of variable of each integral in the sum we get a state identical to \eqref{cstatefinal}.

\subsubsection{The $SU(2)$ example.} For Lie algebras of lower dimension the formula for the generators given in \eqref{goodguys} becomes very simple. In particolar the three generators of $SU(2)$ take the form\footnote{The generators in \eqref{su2gen} differs from the one in \eqref{goodguys} by a normalization factor to follow the conventions we found in the literature. A quotient with respect to the total number operator is needed to reduce $U(2)$ to $SU(2)$.}
\begin{align}
\label{su2gen}
J_+= a^\dagger_{k_1} a_{k_2}\,, \quad
J_-= a^\dagger_{k_2} a_{k_1}\,,\\
J_3= \frac{1}{2}\left(a^\dagger_{k_1} a_{k_1} - a^\dagger_{k_2} a_{k_2}\right)\,.
\end{align}
If we define a generalized coherent state with reference vector the vacuum, we find that
\begin{equation} \nonumber
\exp \left[\int_{k_1\neq k_2} \alpha_1(k_1,k_2) a_{k_1}^\dagger a_{k_2} + \alpha_2(k_1,k_2) a_{k_2}^\dagger a_{k_1}  +  \alpha_3^*(k_1,k_2)\left(a^\dagger_{k_1} a_{k_1} - a^\dagger_{k_2} a_{k_2}\right)  \right] \left| 0 \right>=\left| 0 \right>\,.
\end{equation}
This corresponds to the fact that the vacuum transforms in the trivial representation of $SU(2)$ (and $SU(N)$ in general). Thus it is not possible to build a generalized coherent state for $SU(2)$ while representng the generators of the algebra on bosonic Hilbert space.

\subsubsection{The $SU(2,1)$ example.} Similarly, the eight generators of $SU(2,1)$ take the form
\begin{align}
\label{su21gen}
G_1= -a^\dagger_{k_1} a_{k_2}\,, \qquad
G_2= -a^\dagger_{k_2} a_{k_1}\,,\\
G_3= a^\dagger_{k_1} a_{k_1} - a^\dagger_{k_2} a_{k_2}\,, \\
G_4 = a^\dagger_{k_3} a^\dagger_{k_1}\,, \qquad  G_5 = -a_{k_3} a_{k_1}\,,\\
G_6 =a_{k_3} a_{k_2}\,, \qquad  G_7 = -a_{k_3} a_{k_2}\,,\\
G_8= a_{k_3}^\dagger a_{k_3} + a_{k_1}^\dagger a_{k_1} +1\,.
\end{align}
When we construct a generalized coherent state the only operators that has a non-trivial action on the vacuum are $G_4$ and $G_6$. Modulus a normalization factor,
\begin{equation}
\exp \left[\int_{k_1\neq k_2\neq k_3} \alpha_1(k_1,k_2,k_3) a^\dagger_{k_3} a^\dagger_{k_1} + \alpha_2(k_1,k_2,k_3) a^\dagger_{k_3} a^\dagger_{k_2}  \right] \left| 0 \right> \end{equation}
If we impose than that the state is homogeneous as we did in \eqref{homogeneouseq} we obtain the following sets of solutions for the coeffients $\alpha_1 \propto \delta(k_1+k_3)$ and $\alpha_2=0$ or $\alpha_2 \propto \delta(k_2+k_3)$ and $\alpha_1=0$. In both cases the coherent state reduces trivially to a $SU(1,1)$ coherent state.

\subsubsection{The $SU(2,2)$ example.} The last case we investigate explicitly is $SU(2,2)$ its fifteen  generators take the form
 \begin{align}
\label{su22gen}
G_1= -a^\dagger_{k_1} a_{k_2} \quad
G_2= -a^\dagger_{k_2} a_{k_1}\\
G_3= a^\dagger_{k_1} a_{k_1} - a^\dagger_{k_2} a_{k_2} \\
G_4= a^\dagger_{k_3} a_{k_4} \quad
G_5= a^\dagger_{k_4} a_{k_3}\\
G_6= a^\dagger_{k_1} a_{k_1} + a^\dagger_{k_3} a_{k_3} +1\\
G_7= a^\dagger_{k_1} a_{k_1} + a^\dagger_{k_4} a_{k_4} +1\\
G_8= a^\dagger_{k_1} a^\dagger_{k_3}  \quad G_9= -a_{k_1} a_{k_3} \\
G_{10}= a^\dagger_{k_1} a^\dagger_{k_4}  \quad G_{11}= -a_{k_1} a_{k_4}\\
G_{12}= a^\dagger_{k_2} a^\dagger_{k_3}  \quad G_{13}= -a_{k_2} a_{k_3} \\
G_{14}= a^\dagger_{k_2} a^\dagger_{k_4}  \quad G_{15}= -a_{k_2} a_{k_4} \\
\end{align}
In analogy with the $SU(2,1)$ construction the only non trivial contribution to the coherent state comes from the action on the vacuum of the generators $G_8$, $G_{10}$, $G_{12}$, $G_{14}$. Modulus a normalization factor,
\begin{align}
\exp \Bigg[ \int_{k_1\neq k_2 \neq k_3 \neq k_4} &\alpha_1(k_1,k_2,k_3,k_4) a^\dagger_{k_1} a^\dagger_{k_3} + \alpha_2(k_1,k_2,k_3,k_4) a^\dagger_{k_2} a^\dagger_{k_3} \\
& + \alpha_3(k_1,k_2,k_3,k_4) a^\dagger_{k_1} a^\dagger_{k_4}   + \alpha_4(k_1,k_2,k_3,k_4) a^\dagger_{k_2} a^\dagger_{k_4}  \Bigg] \left| 0 \right>
\end{align}
If we impose than that the state is homogeneous as we did in \eqref{homogeneouseq} we obtain the following sets of non-trivial solutions $\alpha_2= \alpha_3 = 0$, $\alpha_1 \propto \delta(k_1+k_3)$, $\alpha_4 \propto \delta(k_2+k_4)$ or $\alpha_1= \alpha_4 = 0$, $\alpha_2 \propto \delta(k_1+k_4)$, $\alpha_3 \propto \delta(k_2+k_3)$. Let's consider the first case for example
\begin{align}
\exp \Bigg[ \int_{k_1\neq k_2 \neq k_3 \neq k_4} \hspace{-2em} \alpha_1(k_1,k_2,k_3,k_4) a^\dagger_{k_1} a^\dagger_{-k_1}   + \alpha_4(k_1,k_2,k_3,k_4) a^\dagger_{k_2} a^\dagger_{-k_2}  \Bigg] \left| 0 \right> =\\
\exp \Bigg[ \int_{k_1\neq k_2 \neq k_4} \hspace{-2em}\alpha_1(k_1,k_2,-k_1,k_4) a^\dagger_{k_1} a^\dagger_{-k_1}\Bigg]   \exp \Bigg[ \int_{k_1\neq k_2 \neq k_3 } \hspace{-2em} \alpha_4(k_1,k_2,k_3,-k_2) a^\dagger_{k_2} a^\dagger_{-k_2}  \Bigg] \left| 0 \right> =\\
\exp \Bigg[ \int_{k} \tilde\alpha_1(k) a^\dagger_{k} a^\dagger_{-k}\Bigg]   \exp \Bigg[ \int_{k} \tilde\alpha_4(k) a^\dagger_{k} a^\dagger_{-k}  \Bigg] \left| 0 \right> =\\
\exp \Bigg[ \int_{k} \left(\tilde\alpha_1(k)+\tilde\alpha_4(k)\right) a^\dagger_{k} a^\dagger_{-k}\Bigg] \left| 0 \right>
\end{align}
where in the first line we used that the operator $a^\dagger_{k_1} a^\dagger_{-k_1}$ commutes with $a^\dagger_{k_2} a^\dagger_{-k_2}$ so the exponentials factorize, in the second line we performed formally the integrals over the extra momenta variables, and we renamed the third momenta to $k$, in the last line we get a generic $SU(1,1)$ coherent state. For the  second set of solutions the conclusion is exactly the same.

\subsubsection{Is it possible to generate state with non Gaussianities?}
To get a coherent state that gives a non vanishing expectation value to an odd number of fields we would need a realization of the generators of the algebra, that transforms in the adjoint representation, made by an odd number of ladder operators. From \eqref{bogoliubovop} we can find out that the doublet $(a_k , a^\dagger_{-k} )$ transform in the fundamental representation, and in the tensor product of an odd number of fundamental representations there are no copies of the adjoint, it follows that it is not possible to form generators with an odd number of ladder operators.


\section{Quantum signatures in CMB observables}\label{4}
In this section, we shall use the procedure outlined in \cite{Dona:2016fip}, in order to calculate cosmological perturbations as quantum fluctuations of group coherent states. This shall have profound phenomenological consequences since the cosmological signatures for these states are different from the BD one. For simplicity, we restrict our analysis to the case in which only a real scalar field (playing the role of an inflaton), with quadratic potential, is coupled to gravity. This is the case that is referred to within the literature as chaotic inflation \cite{Linde}.

\subsection{Spectrum of scalar perturbations: chaotic inflation}
Our goal is to calculate the primordial power spectrum for the curvature perturbation for non-BD initial states, in the specific case of chaotic inflation. However, instead of calculating it using the usual formalism of scalar perturbations in Bogolyubov transformed BD modes, we employ fluctuations of coherent states  \cite{Dona:2016fip}. The power spectrum of scalar perturbations shall be recovered while working in the so-called flat gauge.

In slow-roll inflation, the scalar field $\phi$ is slowly rolling toward the bottom of its potential well, here denoted with $V(\phi)$: its kinetic energy $K_\phi=\dot{\phi}^2/2$ is negligible, and its energy density is totally approximated by its potential energy. Since $\phi$ is almost constant, we can set up the initial condition $\phi_0$, which also fixes the value of the potential. The relevant equations are then provided by the approximation on the energy density, and by the approximated equation of motion:
\begin{eqnarray}
\rho\simeq V(\phi)\, \qquad {\rm and} \qquad 3H\dot{\phi}\simeq V'\,.
\end{eqnarray}
The `flatness' which is required of the potential $V(\phi)$ for these approximate relations to be valid, traces back to a requirement on its functional derivatives in $\phi$. This can be estimated introducing the slow-roll parameters for cosmological inflation. The latter are denoted as $\epsilon_{\rm s.r.}$ and $\eta_{\rm s.r.}$, and read
\begin{eqnarray}
\epsilon_{\rm s.r.}\equiv \frac{m^2_{\rm Pl}}{2}\, \left( \frac{V'}{V} \right)^2, \quad {\rm and} \quad \eta_{\rm s.r.}\equiv m^2_{\rm Pl}\, \frac{V''}{V}\,.
\end{eqnarray}

For chaotic inflation, once we redefine the field via $\chi=a(\eta) \phi$, the equation of motion for the background field $\chi$ can be expressed in conformal coordinates $\{\eta, x^i\}$, in which the lone dynamical quantity is given by $ds^2=a(\eta)^2(-d\eta^2+d\vec{x}^2)$, as
\begin{eqnarray} \label{sfr}
\chi''-\nabla^2 \chi + \left( a^2 m^2 \!-\!\frac{a''}{a} \right) \chi =0\,.
\end{eqnarray}
Quantization of the $\chi$ field is attained imposing canonical commutation relations: ladder operators are hence defined, together with the action on the BD vacuum in our case of dS, as explained in the Sec-\ref{2}. The Fourier space-modes $\chi_{\vec{k}}(\eta)$ of the $\chi$ field are then determined by solving the equation
\begin{eqnarray} \label{chiom}
\chi_{\vec{k}}(\eta)''\!+\!(a H)^2 \!\left[  \left( \frac{m}{H} \right)^2 \!+\! \left( \frac{k}{a H} \right)^2 \!\!-\!\frac{H'}{H}\!-\!2 \right] \!\chi_{\vec k}(\eta) = 0,\nonumber
\end{eqnarray}
in which we have used $a''/a^3=H'+2H^2$. A solution to \eqref{chiom} can be recovered on de Sitter background, while matching it to the Bunch-Davies vacuum \cite{Bunch:1978yq}. In terms of Hankel functions, Fourier mode functions read
\begin{eqnarray}
\chi_{\vec{k}}(\eta)=\sqrt{ \frac{-\eta \pi}{2}} \, e^{\imath \frac{\pi}{4} (2 \nu +1) }\, H_\nu^{(1)} (-k \eta)\,,
\end{eqnarray}
in which the label $\nu$ show a dependence on the parameters of the theory, i.e.
\begin{eqnarray}
\nu=\sqrt{\frac{9}{4}-\left( \frac{m}{H}\right)^2}\,.
\end{eqnarray}
During the dS phase, the universe evolves with a scale factor given by $\eta=-(a H)^{-1}$. For the super-horizon modes that satisfy $|k \eta|\ll 1$, or in other words, after crossing the Hubble horizon, the solution evolves as
\begin{eqnarray} \label{sol}
\chi_{\vec{k}}(\eta)= \frac{e^{\imath \frac{\pi}{4} (2\nu -1) }}{\sqrt{2 k }}\, \frac{\Gamma(\nu)}{\sqrt{\pi}}\, \left( \frac{-k \eta}{2}\right)^{\frac{1}{2}-\nu}\,.
\end{eqnarray}

Taking into account that for slow-roll inflation the expressions $\rho\simeq V(\phi)$ and $3 H \dot{\phi}\simeq - V'(\phi)$ hold, the curvature perturbation variable can be recovered to be
\begin{eqnarray}
\zeta=\frac{1}{3}\frac{V'}{\dot{\phi}^2}\delta \phi = \frac{1}{m_{\rm Pl}^2}\frac{V}{V'} \delta \phi\,,
\end{eqnarray}
in which linear perturbation of the scalar field are defined following the recipe in \cite{Dona:2016fip}. We shall use that prescription to define the cosmological perturbations about a   non-BD state. Given a coherent state $|\alpha\rangle=\prod_{\vec{k}} |\alpha_{\vec{k}}\rangle$, for which the perturbation of the mode functions $\alpha'=\alpha+ \delta \alpha+ \cdots$ is considered, it holds
\begin{eqnarray}
\delta\phi= \langle \alpha+ \delta \alpha| \phi | \alpha+ \delta \alpha \rangle \Big|_{O(\delta \alpha)}\,.
\end{eqnarray}

The power spectrum can finally be computed from the perturbed expectation value of the correlation function of the $\chi$ field, once we expand at the second order the quantity
\begin{eqnarray}
&&\!\!\!\!\!\! \langle  \alpha+ \delta \alpha | \hat{\chi}(\eta, \vec{x}) \hat{\chi}(\eta, \vec{y}) | \alpha+ \delta \alpha \rangle\Big|_{O(\delta \alpha^2)} =\\
&&\!\!\!\!=\frac{1}{2\pi^2} \int_0^\infty\!\!\!\! k^3\, |\delta \alpha_{\vec{k}}|^2 \, |\chi_{\vec{k}}(\eta)|^2\, \frac{\sin kL}{k L}\, \frac{dk}{k}
\equiv \int_0^\infty \!\!\!\! k^3 \mathcal{P}_{\delta \chi}\, \frac{\sin kL}{k L}\, \frac{dk}{k}, \nonumber
\end{eqnarray}
where the coordinate space distance $L=|\vec{x}-\vec{y}|$ has been considered and the power spectrum  $\mathcal{P}_{\delta \chi}= (k^3/2\pi^2)\, |\delta \alpha_{\vec{k}}|^2 \,|\chi_{\vec{k}}|^2$ has been defined. The parameter $L$ can be linked experimentally to a pivot scale, usually referred to as $k_0$.

It is important to note that in order to recover scale-invariance of the power spectrum, it is not sufficient to impose $\nu=3/2$, as it usually happens in the literature when considering the expectation value of the two-point function of the field perturbations in the vacuum state. Instead, the density distribution $|\delta \alpha_{\vec{k}}|^2$ can now provide some extra $k$ factor dependence, which shall then be taken into account. This is the result of having a non-BD initial state, which contains quanta of excited particles in it. If $|\delta \alpha_{\vec{k}}|^2$ is flat in $k$, and $\nu=3/2$, then the power spectrum can be cast in terms of $\mathcal{P}_{\delta\phi}$, by considering that $\delta \phi = \delta \chi/a$. Finally, for $\nu=3/2$ we find
\begin{eqnarray}
\mathcal{P}_{\delta\phi}=\left(\frac{H}{2\pi} \right)^2\,,
\end{eqnarray}
which is indeed scale-invariant. For inflation, the Hubble parameter can be evaluated at the horizon-exit, for modes such that $a H_k=k$, so that in in Planckian units $H\simeq 10^{-5}$, the experimental value \cite{WMAP, Planck} $\mathcal{P}_\zeta(k_0)=(2.445\pm0.096)\times 10^{-9}$ is recovered, having set the pivot scale to $k_0=0.002{\rm Mpc}^{-1}$.

If we allow for this simple choice of the $\nu$ parameter and parametrize the mode density perturbation $|\delta \alpha_{\vec{k}}|^2\sim k^{n_\alpha}$, deviations from scale invariance are attained, for a light scalar field mass $m\leq 3H/2$ such that
\begin{eqnarray}
\nu \simeq \frac{3}{2} -\frac{m^2}{3 H^2}\,,
\end{eqnarray}
as it follows
\begin{eqnarray} \label{uno}
\mathcal{P}_{\delta\phi}=\left(\frac{H}{2\pi} \right)^2\, \left(\frac{k}{2 a H} \right)^{\frac{2}{3} (\frac{m}{H})^2 + n_\alpha}\,,
\end{eqnarray}
This extra polynomial dependence now arising from $|\delta \alpha_{\vec{k}}|^2$ now enters the expression for the spectral index, which is defined as
\begin{eqnarray}
n-1\equiv \frac{d \ln \mathcal{P}}{d \ln k}\,.
\end{eqnarray}

Recalling that for almost scale invariant massive fields, for which $\nu=3/2$, it holds
\begin{eqnarray} \label{due}
\mathcal{P}_{\zeta}= \left(  \frac{1}{m_{\rm Pl}^2}\frac{V}{V'}  \right)^2 \!\!\left( \frac{H_k}{2\pi}\right)^2\!\!\simeq \frac{1}{24 m_{\rm Pl}^4 \pi^2 }\,\frac{V}{\epsilon_{\rm s.r.}}\,,
\end{eqnarray}
which is evaluated at the horizon exit, we can finally cast the power spectrum as
\begin{eqnarray}
\mathcal{P}_\zeta(k)=\mathcal{P}_\zeta(k_0)\left( \frac{k}{k_0}\right)^{n-1}\,.
\end{eqnarray}
Within the assumption that the spectral index $n$ has a vanishing running parameter $\alpha$, the best fit for experiments \cite{WMAP, Planck} is found to be
\begin{eqnarray} \label{tre}
n=0.960\pm0.013\,.
\end{eqnarray}

How does all this tie up with the usual description of non-BD states (which does \textit{not} consider them as fluctuations of a coherent state)? To answer this question, we only need to recall the form of the power spectrum for such states, as given in, say, \cite{Ganc}
\bea
P_\zeta(k) = P_\zeta^{\text{BD}}(k) |\alpha_k + \beta_k|^2\,.
\eea
The extra boost in the power spectrum in that case $|\alpha_k + \beta_k|^2$, written in terms of the Bogolyubov coefficients, are simply the $|\delta\alpha_{\vec{k}}|^2$ in our case\footnote{We apologize for the slightly confusing choice of notation for our coherent states. A priori the $\delta\alpha$ has no relation with the Bogolyubov coefficient $\alpha$.}. These factors have been constrained greatly by using the both backreaction of the excited modes as well as running of the spectrum (from \eqref{uno}, \eqref{due} and \eqref{tre}). Thus, signatures of the coherent state can not only be seen in the power spectrum, but also has been tightly constrained by data so far.

\subsection{Primordial non-Gaussianities}
Having extracted as much as one can from the power spectrum, it s important to turn to the primordial non-Gaussianities of inflationary perturbations. As we shall demonstrate quite easily, these also act as important probes for signatures of coherent states in the cosmological data. In particular, we shall only focus on the leading order of the non-linearity parameter, the bispectrum, to argue our case. Even for single field models of inflation, we can have different shapes as well as amplitudes of the bispectrum depending on the details of the theory. For instance, it is well known that a bispectrum peaked in the equilateral shape can arise from a model of inflation with a `small' speed of sound $c_s$. Such a model depends on having higher order kinetic terms in the action and results in the spatial derivatives and the time derivatives evolving with different coefficients (DBI inflation being a prime example of this). On the other hand, \cite{Mald} had establised that for vanilla slow roll models of inflation, such as our example of the chaotic potential, the $f_{NL}$ is very small, being proportional to the slow-roll parameters, $\epsilon_{\rm s.r.}$ and $\eta_{\rm s.r.}$, in the comoving gauge. In fact, it has been shown that the scalar bispectrum of this kind of simple models of inflation, in the squeezed limit, do not have any observable effects of non-Gaussianity for our universe, when working in suitable Fermi normal coordinates. However, this does imply that observing a positive result for $f_{NL}$ shall rule out all such models, including our particular case of chaotic inflation. This is where the effect of having a non-BD state shall play a crucial role in our argument.

As shown in the previous sub-section, the fluctuation of our coherent state, parametrized as $|\delta\alpha_{\vec{k}}|^2$, is essentially the same as having a Bogolyubov rotated BD state. For completeness, we quote the result of the exact calculation of the bispectrum for this non-BD state.
\bea
& &B_\zeta(k_1, k_2, k_3)\approx\frac{1}{8} \frac{H^4}{\epsilon_{\rm s.r.}} \frac{1}{k_1k_2k_3} \sum_i \left(\frac{1}{k^2_i}\right) (\alpha_{k_1} + \beta_{k_1}) (\alpha_{k_2} + \beta_{k_2}) (\alpha_{k_3} + \beta_{k_3})\times\\
& & \;\;\;\left[\frac{1}{k_1 + k_2 + k_3}(\bar{\alpha}_{k_1}\bar{\alpha}_{k_2}\bar{\alpha}_{k_3} - \bar{\beta}_{k_1}\bar{\beta}_{k_2}\bar{\beta}_{k_3}) +  \frac{1}{k_1 + k_2 - k_3} (\bar{\alpha}_{k_1}\bar{\alpha}_{k_2}\bar{\beta}_{k_3} - \bar{\beta}_{k_1}\bar{\beta}_{k_2}\bar{\alpha}_{k_3}) \right.\nn\\
& & \;\;\;\;\;\left. \frac{1}{k_1 - k_2 +k_3} (\bar{\alpha}_{k_1}\bar{\beta}_{k_2}\bar{\alpha}_{k_3} - \bar{\beta}_{k_1}\bar{\alpha}_{k_2}\bar{\beta}_{k_3}) + \frac{1}{k_1 - k_2 - k_3}(\bar{\alpha}_{k_1}\bar{\alpha}_{k_2}\bar{\beta}_{k_3} - \bar{\beta}_{k_1}\bar{\alpha}_{k_2}\bar{\alpha}_{k_3})\right] +\text{c.c.}\nn
\eea
In the above, an overbar denotes complex conjugation. It can be shown that this bispectrum actually peaks in the folded configuration, when $k_1 \simeq 2k_2 \simeq 2k_3$, and can thus be observable. This leads to the well known folded shape of the bispectrum, in the case of a non-BD vacuum. The folded-type non-Gaussianity can be parametrized by a form of the bispectrum given by
\bea
B_\zeta^{\text{folded}} (k_1, k_2, k_3) &=& \frac{3}{5} f_{NL}^{\text{folded}} \left[-\frac{3}{k_1^{4-n_s} k_2^{(4-n_s)}} -\frac{3}{k_1^{(2-n_s)/3} k_2^{(2-n_s)/3}k_3^{(2-n_s)/3}} \right.\nn\\
&& \;\;\left.+\frac{3}{k_1^{(2-n_s)/3} k_2^{(2-n_s)/3}k_3^{(2-n_s)/3}} + \text{perms.}\right]\,.
\eea
This can also be written as a combination of the equilateral and the orthogonal shapes. Thus a detection of a folded-type non-Gaussianity, or even an absence of it, shall be able to put even more stricter forms on the functional form of $|\delta\alpha_{\vec{k}}|^2$, and thus on the coherent state.

\section{Conclusions}\label{5}
The investigation of non-BD states is particularly fruitful while enlightening rich phenomenological consequences for both late time and early time cosmological observables. Although in this study we have mainly focused on the richness that this scenario may occasion for what concerns inflation\footnote{At this purpose a detailed study of the phenomenological consequences of using bosonic non Bunch-Davies states might be fruitfully devoted to models of vector inflation \cite{Ford:1989me, Parker:1993ya, Golovnev:2008cf, Maleknejad:2011sq, Alexander:2011hz, Adshead:2012kp, Alexander:2014uza}.}, is nonetheless true that quantum features proper of many-body systems may also affect measurements at small redshift, including entanglement harvesting and many other redolent alternatives recently elaborated in experimental cosmology.

From a theoretical point of view, the emergence of non Bunch Davies vacuum states in cosmology is closely related to the incontrovertible existence of thermal properties for the universe, and to their evolution. What actually encompasses these two concepts is the possibility of recovering by means of Bogolyubov transformations a Gibbs state, showing thermal properties for both bosonic or fermionic statistics, from a pure vacuum state. Whenever the vacuum state coincides with the one recovered on a dS background, this will be denoted as a Bunch Davies vacuum. The thermal state arising from applying a Bogolyubov transformation to the Bunch Davies vacuum, is in stead called non Bunch Davies, and represents a quantum state with thermal modes distribution. The perturbation of their modes densities characterize the cosmological perturbations, entering in the relevant CMB observables.

Indeed, we pointed out that important modifications of physical observables arise while non Bunch Davies vacuum states are taken into account, specifically in the spectrum of scalar perturbations, and in the generation of non-gaussianities. Thus, the framework we have deepened here  must be considered in its wide phenomenological novelty. Nonetheless, other quantum features entering cosmological observables might be accessible in the future, most notably quantum entanglement of modes which results from the expansion of the Universe. Although this scenario seems to be unlikely to become falsifiable for early time cosmology, at least as soon as CMB measurements are taken into account, (unexpected) improvements on the detection of neutrinos cosmological background might represent a further source of attention to this proposal. Anyway, more substantial chances can arise for early time cosmology, from entanglement-harvesting at small redshift, and the eventual confirmation of the expansion of the universe from purely quantum effects. These represent all fascinating topics to which we foresee that our proposal of generalized non Bunch Davies states may contribute through forthcoming investigations.

\section*{Acknowledgements}
We acknowledge support by the Shanghai Municipality, through the grant No. KBH1512299, and by Fudan University, through the grant No. JJH1512105.


\begin{thebibliography}{99}

\bibitem{Birrell:1982ix}
  N.D.~Birrell and P.C.W.~Davies,
  ``Quantum Fields in Curved Space,''
  doi:10.1017/CBO9780511622632 .


\bibitem{Bunch:1978yq}
  T.~S.~Bunch and P.~C.~W.~Davies,
  ``Quantum Field Theory in de Sitter Space: Renormalization by Point Splitting,''
  Proc.\ Roy.\ Soc.\ Lond.\ A {\bf 360}, 117 (1978).

\bibitem{Ion}
 I.~V.~Vancea,
 ``Entanglement Entropy in the $\sigma$-Model with the de Sitter Target Space,''
  [arXiv:1609.02223 [hep-th]].

\bibitem{Kolb}
E.~W.~Kolb and M.~S.~Turner,
``The Early Universe'' (Addison-Wesley, New York, 1990).

\bibitem{Vilenkin:1987kf}
  A.~Vilenkin,
  ``Quantum Cosmology and the Initial State of the Universe,''
  Phys.\ Rev.\ D {\bf 37}, 888 (1988).
  doi:10.1103/PhysRevD.37.888

\bibitem{Brandenberger:2000wr}
  R.~H.~Brandenberger and J.~Martin,
  ``The Robustness of inflation to changes in superPlanck scale physics,''
  Mod.\ Phys.\ Lett.\ A {\bf 16}, 999 (2001)
  doi:10.1142/S0217732301004170
  [astro-ph/0005432].

\bibitem{Kaloper:2002cs}
  N.~Kaloper, M.~Kleban, A.~Lawrence, S.~Shenker and L.~Susskind,
  ``Initial conditions for inflation,''
  JHEP {\bf 0211}, 037 (2002)
  doi:10.1088/1126-6708/2002/11/037
  [hep-th/0209231].

\bibitem{Holman:2007na}
  R.~Holman and A.~J.~Tolley,
  ``Enhanced Non-Gaussianity from Excited Initial States,''
  JCAP {\bf 0805}, 001 (2008)
  doi:10.1088/1475-7516/2008/05/001
  [arXiv:0710.1302 [hep-th]].

\bibitem{Ashoorioon:2010xg}
  A.~Ashoorioon and G.~Shiu,
  ``A Note on Calm Excited States of Inflation,''
  JCAP {\bf 1103}, 025 (2011)
  doi:10.1088/1475-7516/2011/03/025
  [arXiv:1012.3392 [astro-ph.CO]].

\bibitem{Ashoorioon:2013eia}
  A.~Ashoorioon, K.~Dimopoulos, M.~M.~Sheikh-Jabbari and G.~Shiu,
  ``Reconciliation of High Energy Scale Models of Inflation with Planck,''
  JCAP {\bf 1402}, 025 (2014)
  doi:10.1088/1475-7516/2014/02/025
  [arXiv:1306.4914 [hep-th]].
  
\bibitem{Ashish}
 A.~Shukla, S.~P.~Trivedi and V.~Vishal,
 ``Symmetry constraints in inflation, $\alpha$-vacua, and the three point function,''
 JHEP {\bf 1612}, 102 (2016)
 doi:10.1007/JHEP12(2016)102 
 [arXiv:1607.08636 [hep-th]].

\bibitem{Benedict}  
 B.~J.~Broy,
 ``Corrections to $n_s$ and $n_t$ from high scale physics,''
 Phys.\ Rev.\ {\bf D94}, 103508 (2016)
 doi:10.1103/PhysRevD.94.109901, 10.1103/PhysRevD.94.103508
 [arXiv:1609.03570 [hep-th]].

\bibitem{Dona:2016fip}
  P.~Don\`a and A.~Marcian\`o,
  ``A note on the semiclassicality of cosmological perturbations,''
  accepted in Phys. Rev. D, arXiv:1605.09337 [gr-qc].

\bibitem{WiMi}
H.~ M.~Wiseman and G.~J.~Milburn, ``Quantum Measurement and Control'', (Cambridge University Press, 2009).

\bibitem{MartinMartinez:2012sg}
  E.~Martin-Martinez and N.~C.~Menicucci,
  ``Cosmological quantum entanglement,''
  Class.\ Quant.\ Grav.\  {\bf 29}, 224003 (2012)
  [arXiv:1204.4918 [gr-qc]].

\bibitem{Martin-Martinez:2014gra}
  E.~Martin-Martinez and N.~C.~Menicucci,
  ``Entanglement in curved spacetimes and cosmology,''
  Class.\ Quant.\ Grav.\  {\bf 31}, no. 21, 214001 (2014)
  [arXiv:1408.3420 [quant-ph]].


\bibitem{R1}
B.~Reznik, ``Entanglement from the Vacuum," Found. Phys. V33, 167 (2003).

\bibitem{R2}
B.~Reznik, A.~Retzker, and J.~Silman, ``Violating Bell's inequalities in vacuum,'' Phys. Rev. A 71, 042104 (2005).

\bibitem{VM}
G.~Ver~Steeg and N.~C.~Menicucci, ``Entangling Power of an Expanding Universe,'' Phys. Rev. D 79, 044027 (2009).

\bibitem{N}
Y.~Nambu, ``Entanglement Structure in Expanding Universes,'' Entropy 15, 1847 (2013).


\bibitem{Had}  J.~Hadamard, {\it ``Lectures on CauchyÕs Problem in Linear Partial Differential Equations} (Yale University Press, New Haven, CT, 1923); F.~G.~Friedlander, {\it ``The Wave Equation on a Curved Space-Time''} (Cambridge University Press, Cambridge, 1975).

\bibitem{Perelomov} A.~Perelomov, Generalized coherent states and their applications, Springer, Berlin 1986.

\bibitem{Schwinger} J.~Schwinger, 1952 Atomic Energy Commission Report No NYO-3071.

\bibitem{Barut} A.~O.~Barut, C.~Fronsdal, Proc. R. Soc. Lond. A 1965 287 532-548.

\bibitem{Anderson} R.~Raczka, J.~Fischer, and R.~L.~Anderson, Proc. Royal Soc. of London. Series A, Vol. 302 1471, pp. 409-500.

\bibitem{VilenkinFord} A.~Vilenkin and L.~Ford, Gravitational effects upon cosmological phase transitions, Phys. Rev. D 26, 1231, 1982.

\bibitem{Nishant} N.~Agarwal, R.~Holman, A.~Tolley and J.~Lin, Effective field theory and non-Gaussianity from general inflationary states, JHEP 1305 (2013) 085.

\bibitem{Maldacena} J.~Maldacena, ``Non-Gaussian features of primordial fluctuations in single field inflationary models,'' JHEP 0305 (2003) 013.

\bibitem{Ganc} J.~Ganc, ``Calculating the local-type fNL for slow-roll inflation with a non-vacuum initial state,'' 	Phys. Rev. D 84, 063514 (2011).

\bibitem{Linde} A.~D.~Linde,  ``Chaotic inflation," Phys. Lett. B {\bf 129} (3), 171.

\bibitem{WMAP} WMAP Collaboration, The Astro. Jour. Sup. {\bf 208}, 19, 25 pp. (2013); WMAP Collaboration, The Astro. Jour. Sup. {\bf 208}, 20, 54 pp. (2013)

\bibitem{Planck} Planck Collaboration,``Planck 2015 results. XI. CMB power spectra, likelihoods, and robustness of parameters,'' Submitted to: Astron.Astrophys. arXiv:1507.02704.

\bibitem{ABMR}
V. Acquaviva, N. Bartolo, S. Matarrese and A. Riotto, Nucl.Phys. B667,  119-148 (2003).

\bibitem{Mald}
J. Maldacena, JHEP 0305, 013 (2003).

\bibitem{PajerSchmidt}
E. Pajer, F. Schmidt and M. Zaldarriaga, Phys. Rev. D 88, 083502 (2013).

\bibitem{Ford:1989me}
  L.~H.~Ford,
  ``Inflation Driven By A Vector Field,''
  Phys.\ Rev.\ D {\bf 40}, 967 (1989).
  doi:10.1103/PhysRevD.40.967

\bibitem{Parker:1993ya}
  L.~Parker and Y.~Zhang,
  ``Relativistic condensate as a source for inflation,''
  Phys.\ Rev.\ D {\bf 47}, 416 (1993).
  doi:10.1103/PhysRevD.47.416

\bibitem{Golovnev:2008cf}
  A.~Golovnev, V.~Mukhanov and V.~Vanchurin,
  ``Vector Inflation,''
  JCAP {\bf 0806}, 009 (2008)
  doi:10.1088/1475-7516/2008/06/009
  [arXiv:0802.2068 [astro-ph]].

\bibitem{Maleknejad:2011sq}
  A.~Maleknejad and M.~M.~Sheikh-Jabbari,
  Phys.\ Rev.\ D {\bf 84}, 043515 (2011)
  doi:10.1103/PhysRevD.84.043515
  [arXiv:1102.1932 [hep-ph]].

\bibitem{Alexander:2011hz}
  S.~Alexander, A.~Marciano and D.~Spergel,
  ``Chern-Simons Inflation and Baryogenesis,''
  JCAP {\bf 1304}, 046 (2013)
  doi:10.1088/1475-7516/2013/04/046
  [arXiv:1107.0318 [hep-th]].

\bibitem{Adshead:2012kp}
  P.~Adshead and M.~Wyman,
  ``Chromo-Natural Inflation: Natural inflation on a steep potential with classical non-Abelian gauge fields,''
  Phys.\ Rev.\ Lett.\  {\bf 108}, 261302 (2012)
  doi:10.1103/PhysRevLett.108.261302
  [arXiv:1202.2366 [hep-th]].

\bibitem{Alexander:2014uza}
  S.~Alexander, D.~Jyoti, A.~Kosowsky and A.~Marciano,
  ``Dynamics of Gauge Field Inflation,''
  JCAP {\bf 1505}, 005 (2015)
  doi:10.1088/1475-7516/2015/05/005
  [arXiv:1408.4118 [hep-th]].



\end{thebibliography}
\end{document}